\documentclass[final]{hubsarxiv}

\usepackage{tabularx}
\usepackage{longtable}
\usepackage{textcomp}

\title{A Comprehensive Evaluation Framework for Conversational Home Energy Management Systems}
\shorttitle{Evaluation framework for conversational HEMS}
\shortauthors{Jung}
\author{Wooyoung Jung\affmark{a,}\corrmark\orcid{0000-0003-2746-4983}}
\affil{\affmark{a}Department of Civil and Architectural Engineering and Mechanics, University of Arizona, Tucson, AZ, USA}
\corresponding{wooyoung@arizona.edu}
\paperno{2026-006}
\paperdate{September 2026}
\versionnote{v1 (2026-09-04): first release. v2 (2026-11-05): re-set in the HUBS Lab paper layout; content unchanged.}
\license{CC BY-NC-ND 4.0}
\abstract{The growing complexity in home energy management (HEM) demands advanced systems that guide occupants toward
informed energy decisions while reflecting their background, preferences, and contextual information. Large
language model (LLM)-integrated HEM systems (HEMS) have demonstrated promise, but previous studies relied
on single-turn or single-task evaluations with response accuracy as the primary metric. Consequently,
whether such systems can deliver effective interactions across the extended multi-turn dialogues
typical of real-world use remains an open question. This study introduces a comprehensive evaluation
framework of LLM-integrated HEMS derived from the Goal-Question-Metric methodology,
organized across five categories: task performance, factual accuracy, interaction quality, control capability,
and system efficiency. A total of 23 metrics across multi-turn conversations are proposed and an LLM-as-judge
pipeline is employed to enable scalable automated scoring. Its reliability is
validated against three trained human coders: after iterative rubric calibration, twelve of the
fifteen LLM-scored metrics reached strong agreement (ICC $\geq$ 0.73), three of them perfect, while the
remaining three exhibited near-zero variance in human scores and are instead reported via mean
absolute error (0.04--0.28). To demonstrate the framework's effectiveness, a total of 970
dialogues -- 16 scenarios and five personas -- were generated and evaluated using four conversational HEMS
configurations: a HEMS using a vanilla LLM with raw energy data, one with a vanilla LLM with an energy analysis
report, one with a vanilla LLM with an energy analysis report and chain-of-thought prompting, and a multi-agent HEMS.
The proposed framework successfully distinguished the performance of the four HEMS across multiple evaluation dimensions,
revealing their respective strengths and weaknesses. This study contributes to the advancement of conversational
HEMS by providing a reproducible, multi-dimensional evaluation methodology that comprehensively assesses
sustained, context-aware system performance.}

\begin{document}
\maketitle

\section{Introduction}
\label{sec:introduction}

The energy demand profiles of the residential building sector have become increasingly complex \cite{jin2017foresee} due 
to the multi-functionality of homes \cite{garrote2021who,daniel2020education,amiribesheli2015review}, the widespread 
adoption of distributed energy sources (e.g., rooftop solar photovoltaics) \cite{doe2025solar}, the proliferation of 
high-power end-use technologies such as electric vehicle (EV) chargers \cite{spglobal2025ev}, and growing unpredictability 
in climate. This growing complexity demands advanced home energy management systems (HEMS) capable of coordinating 
heterogeneous energy assets, anticipating dynamic operating conditions, and mediating competing objectives across 
multiple stakeholders \cite{zhou2016smart}. For individual households, such systems are essential for optimizing energy 
costs, maintaining occupant comfort, enabling meaningful participation in demand response (DR) events and self-consumption 
strategies, and even educating occupants about their own energy-use patterns, trade-offs, and the consequences of everyday 
decisions. From a utility and grid-operator perspective, advanced HEMS play a vital role in mitigating peak demand, 
reducing load volatility, and enhancing grid reliability by enabling fine-grained, scalable, and automated control at 
the residential edge.

However, current home energy management (HEM) practices remain largely limited in scope and capability \cite{he2024llm}. 
Most commercial HEMS rely on centralized optimization, static rule-based logic, or single-agent control schemes embedded 
in smart thermostats, in-home displays, or utility-facing DR platforms \cite{hannan2018internet}. Even though these 
approaches have enabled basic automation and incremental efficiency gains \cite{zhou2016smart}, they typically conceptualize 
occupants as passive recipients of control actions, offering limited support for sustained interaction, explicit 
preference elicitation, or interpretability of system decisions. Consequently, existing HEMSs lack mechanisms for 
continuous user feedback and engagement, constraining their ability to learn from occupants over time to deliver truly 
personalized, context-aware HEM \cite{badar2022smart}.

Integrating large language models (LLMs) -- a class of artificial intelligence (AI) models for natural language
understanding and generation \cite{naveed2025comprehensive} -- into HEMS (i.e., conversational HEMS) has 
shown promise in addressing this long-standing limitation \cite{he2024llm,jung2026cot,he2025context,rey2024leveraging,
makroum2025agentic,michelon2025llm}. In particular, LLMs have been shown to encode and reason about household 
characteristics within their decision-making processes, enabling system responses and control actions to be adaptively 
tailored to user contexts. Beyond improving decision-support quality, conversational HEMS also enhanced natural 
language-based interactions with occupants, thereby supporting more transparent, interpretable, adaptive, and 
human-centered energy management. However, existing conversational HEMS remain limited in their performance 
evaluation \cite{he2024llm,jung2026cot,he2025context,rey2024leveraging,makroum2025agentic,michelon2025llm}. 
Most studies primarily relied on accuracy-oriented metrics (i.e., whether the system's response or control 
action was correct) in single-turn or single-task interactions, which do not capture the sustained, multi-turn 
conversational dynamics between occupants and HEMS. When conversational HEMS are deployed in reality, occupants 
are likely to interact with these systems continuously for purposes ranging from basic energy inquiries to 
device control across varying operating conditions \cite{li2025beyond,jung2026humanai}. Hence, when such 
systems are assessed, various evaluation angles are needed to comprehensively understand their strengths 
and weaknesses. This study aims to address this gap and is guided by the following research questions (RQs):

\begin{enumerate}
    \item What evaluation dimensions are necessary for comprehensively assessing conversational HEMS?
    \item Can the proposed evaluation framework employ an LLM-as-judge approach for its scalability?
    \item Can the proposed evaluation framework successfully distinguish different conversational HEMS' performance?
\end{enumerate}

By addressing these questions, this study makes three major contributions: (1)~it proposes and validates a
comprehensive evaluation framework, grounded in the Goal-Question-Metric (GQM) methodology, comprising 23
metrics across five categories -- task performance, factual accuracy, interaction quality, control capability,
and system efficiency -- for assessing conversational HEMS (RQ1); (2)~it employs an LLM-as-judge approach for
scalable automated scoring validated against human coders (RQ2); and (3)~it applies this framework to compare
four HEMS configurations across diverse scenarios and user personas, providing evidence of discriminative
validity (RQ3).


\section{Related Work}
\label{sec:related_work}

\textbf{Conversational home/building energy management systems}: Research efforts in this category designed
LLM-integrated energy management systems to support occupants or building managers in energy-related decision-making.
LLMs have demonstrated broad applicability to smart home intelligence tasks beyond energy management, including
zero-shot activity recognition from sensor streams \cite{civitarese2025llm}, further motivating their integration
into conversational HEMS. These systems differ in their architectural approaches, functional scopes, and evaluation strategies, as summarized in
Table~\ref{tab:literature_review}.

\begin{table}
\centering
\caption{Comparison of existing LLM-integrated HEMS}
\label{tab:literature_review}
\footnotesize
\renewcommand{\arraystretch}{1.2}
\begin{tabularx}{\textwidth}{p{2cm}XXXXX}
\toprule
 & \cite{jung2026cot} & \cite{rey2024leveraging} & \cite{he2024llm,he2025context} & \cite{michelon2025llm} & 
 \cite{makroum2025agentic} \\
\midrule 
System design & Prompt eng.\ (CoT) & Prompt eng.\ + RAG & OpenAI Assistants API & ReAct + few-shot & 
Multi-agent (orchestrator + specialists) \\ 
LLM & GPT-4o-mini & Starling 7B, Qwen 75B & GPT-4o & Mistral 7B & Llama 70B, Qwen 32B, GPT-OSS 120B \\
Functional scope & Energy-saving strategies & Smart home actions & Energy analysis, device control & 
Parameter extraction & Appliance scheduling \\
Interaction type & Single-turn & Single-turn & Single-turn & Agent-driven sequential conversation & 
Single-task \\
User representation & Researcher-designed prompts & Researcher-designed prompts & Researcher-designed prompts & 
LLM-as-user with three difficulty levels & Researcher-designed scenarios \\
Evaluation metrics & Accuracy & User preference reflection in system response, latency & Response accuracy, 
system performance\textsuperscript{*} & Parameter retrieval accuracy & Success rate, system performance\textsuperscript{*} \\
\bottomrule
\multicolumn{6}{l}{\textsuperscript{*}Including system latency, token usage} \\
\multicolumn{6}{l}{Eng.: engineering, RAG: retrieval augmented generation, ReAct: reasoning and acting}
\end{tabularx}
\renewcommand{\arraystretch}{1.0}
\end{table}

As summarized in Table~\ref{tab:literature_review}, these systems span a range of LLM techniques from CoT prompting
\cite{jung2026cot} and RAG-augmented recommendation \cite{rey2024leveraging} to multi-turn ReAct interfaces
\cite{michelon2025llm} and hierarchical multi-agent scheduling \cite{makroum2025agentic}, with functional scopes
ranging from energy analysis and device control \cite{he2024llm,he2025context} to structured parameter extraction
and appliance coordination.

Several evaluation limitations persist across these studies. First, evaluation has relied on narrow, accuracy-oriented
metrics, such as exact parameter match \cite{michelon2025llm}, schedule optimality \cite{makroum2025agentic}, or
intent classification accuracy \cite{rey2024leveraging}, without assessing how systems handle factual
grounding or interaction quality (e.g., whether the system replies to the user's query while considering the
user's characteristics), all of which are critical when occupants interact with conversational HEMS
over time. Second, user representation in evaluation remains limited: most studies used researcher-designed
single-turn queries \cite{jung2026cot,rey2024leveraging,he2024llm,he2025context}, not fully representing
multi-turn dialogue interactions that are more reflective of real-world use. Only one study \cite{michelon2025llm}
used an LLM-as-user approach to simulate multi-turn interactions, but their evaluation focused on parameter retrieval
accuracy like other studies -- i.e., system-oriented metrics that do not capture the quality of human-AI interactions.
Third, cross-domain coordination -- how a system transitions between various user intents
within a single conversation -- remains largely unexplored, despite being relevant to the multi-faceted nature
of real-world HEM interactions.

Notably, only one study examined actual occupant behavior through a human-subject experiment \cite{jung2026humanai},
classifying 156 prompts from 20 participants into six DSS-grounded categories and demonstrating that real
users engage conversational HEMS across multiple functional dimensions simultaneously. This empirical evidence
reinforces that evaluation must account for both interaction quality and system performance across diverse
query types -- a requirement existing frameworks from the broader conversational AI literature only partially address.

\textbf{Evaluation of conversational AI systems}: Prior frameworks assessed dialogue quality through human annotation
\cite{mehri2020usr,mehri2020unsupervised,li2019acute,finch2023abc} or LLM-as-judge approaches
\cite{zheng2023judging,liu2023geval,lin2024wildbench,chang2024survey}. MT-Bench \cite{zheng2023judging} demonstrated that LLMs
can reliably proxy human judges on multi-turn questions; G-Eval \cite{liu2023geval} further showed that
CoT-generated criteria improve alignment with human judgments across diverse tasks.
For task-oriented conversational agents, Wei et al.\ \cite{wei2025enriching} demonstrated that response
quality improves by grounding replies in domain-relevant knowledge -- consistent with the factual grounding
requirements central to HEMS evaluation. Guan et al.\ \cite{guan2026evaluating} further surveyed evaluation
strategies for LLM-based agents in multi-turn conversational settings, covering automated metrics, human
annotation, and LLM-as-judge approaches, but without operationalizing a domain-specific metric set.

Despite these advances, existing frameworks share three limitations for domain-specific HEMS evaluation.
First, they assess open-domain dialogue and do not verify factual grounding against real-world data sources
(e.g., numerical accuracy relative to household energy records). Second, response appropriateness is defined
through generic qualities such as fluency and coherence, not domain-relevant criteria such as calibration
between data-specific and general-knowledge responses. Third, system efficiency -- token usage, latency, and
tool invocation -- is unaddressed despite its relevance for residential deployment.

\section{Evaluation Framework}
\label{sec:framework}

The evaluation of conversational HEMS requires a framework that captures
both functional task performance and the quality of human-AI interactions
expected in HEM.
Accordingly, this study employs the Goal-Question-Metric (GQM) methodology \cite{basili1994goal} to derive
a structured set of evaluation metrics for conversational HEMS. GQM was developed for software process
evaluation \cite{vansolingen1999gqm} and has since been applied to IoT smart home systems \cite{chang2020gqm}
and cloud security \cite{yahya2017gqm}. It is well-suited here because the evaluation goals for
LLM-integrated HEMS are heterogeneous, and existing chatbot evaluation frameworks rely on post-hoc
categorization rather than goal-driven derivation \cite{radziwill2017evaluating,deriu2021survey}. GQM's
top-down structure ensures every metric traces to an explicit goal and every goal is fully operationalized,
providing a principled basis for scalable, automated assessment. This work extends GQM into the evaluation
of multi-turn conversational AI.

GQM proceeds in three steps: (1) define evaluation goals that specify what must be measured; (2) derive operational 
questions that must be answered to assess progress toward each goal; and (3) identify metrics that provide the 
quantitative data needed to answer each question. This top-down derivation ensures that every metric in the 
framework is purposeful and that every goal is operationally measurable. 
To define the evaluation goals (GQM Step 1), three sources were consulted: core HEMS capabilities from
the literature \cite{zhou2016smart,badar2022smart,he2024llm,he2025context,jung2026cot}; an empirical study
of real occupant interactions \cite{jung2026humanai} confirming that multi-dimensional evaluation is
necessary; and the evaluation gaps identified in Section~\ref{sec:related_work}. These sources yielded
five domain-specific evaluation goals and questions:

\begin{itemize}
    \item \textbf{Task Performance}: \textit{Goal} -- Assess the system's ability to accomplish user-requested
    HEM tasks within a multi-turn conversation.
    \textit{Questions} -- Does the system achieve the user's goal before the maximum turn limit? How
    efficiently, in conversational turns, does the system achieve the goal?

    \item \textbf{Factual Accuracy}: \textit{Goal} -- Assess the factual reliability of system responses for
    quantitative claims grounded in household data.
    \textit{Questions} -- What proportion of quantitative claims are accurate relative to ground truth? How
    large is the average error? How many verifiable claims are made, and how many are accurate?

    \item \textbf{Interaction Quality}: \textit{Goal} -- Assess whether system responses and communicative
    behaviors are appropriate to the user's question type, profile, and context, across four dimensions:
    question answering completeness, personalization calibration, technical accessibility, and action
    transparency. \textit{Questions} -- Does the system address user questions? Is personalization matched to
    question type? Does the system make technical content accessible? Does it communicate transparently about
    control actions?

    \item \textbf{Control Capability}: \textit{Goal} -- Assess whether the system executes device control
    actions correctly, including target accuracy, scheduling, mode selection, and constraint compliance.
    \textit{Questions} -- Does the system target the correct device(s)? Are scheduled actions set during
    appropriate time periods? Are device modes appropriate for current conditions? Do actions respect
    device-specific operational limits?

    \item \textbf{System Efficiency}: \textit{Goal} -- Assess computational resource requirements relevant to
    deployment feasibility of LLM-integrated HEMS.
    \textit{Questions} -- How verbose are system responses? How quickly does the system respond? What is the
    total token consumption per conversation?
\end{itemize}

Table~\ref{tab:metrics} operationalizes these goals and questions into 23 measurable metrics. For each metric, 
the table provides the definition and measurement scale. These are measures
that can be objectively extracted from human-AI conversation data; the
evaluation procedure is described in Section~\ref{sec:evaluation}. The number of metrics per category reflects the dimensionality 
of each goal: interaction quality, which spans four sub-dimensions (question answering completeness, personalization 
calibration, technical accessibility, and action transparency), requires finer-grained decomposition than goals 
that address a single evaluative concern.

\begin{small}
\renewcommand{\arraystretch}{1.0}
\begin{longtable}{p{3cm}p{8.5cm}p{2cm}}
\caption{Performance evaluation metrics for LLM-integrated HEMS}
\label{tab:metrics} \\
\toprule
Evaluation metric & Description & Scale \\
\midrule
\endfirsthead
\multicolumn{3}{l}{\small\itshape Table~\ref{tab:metrics} continued from previous page} \\
\toprule
Evaluation metric & Description & Scale \\
\midrule
\endhead
\midrule
\multicolumn{3}{r}{\small\itshape Continued on next page} \\
\endfoot
\bottomrule
\endlastfoot
\multicolumn{3}{l}{\textbf{Task Performance}} \\
Goal achievement rate & Proportion of evaluation runs in which the user's predefined goal was satisfied before 
reaching the maximum turn limit & \% (0--100\%) \\ Turns to completion & Number of conversational turns elapsed 
when the goal was achieved & Count \\
\midrule
\multicolumn{3}{l}{\textbf{Factual Accuracy}} \\
Factual accuracy rate & Proportion of numerical claims that fell within 5\% error of the corresponding ground 
truth value & \% (0--100\%) \\
Mean error percentage & Average absolute percentage error against ground truth across all verifiable numerical 
claims & \% (0--100\%+) \\
Factual claims & Count of verifiable quantitative claims (e.g., energy consumption figures, cost estimates, 
percentages) that can be verified against ground truth data & Count \\
Accurate claims & Count of factual claims verified as accurate ($<$ 5\% error relative to ground truth) & Count \\
\midrule
\multicolumn{3}{l}{\textbf{Interaction Quality}} \\
User questions & Total number of distinct questions posed by the user during the conversation & Count \\
Answered user question ratio & Proportion of user questions that received an appropriate response & \% (0--100\%) \\
Appropriate data-backed response & Count of responses correctly using household data to answer a data-specific 
question & Count \\
Over-personalized response & Count of responses referencing user-specific data to answer a general knowledge 
question not requiring personalization & Count \\
Under-personalized response & Count of responses providing a generic answer to a data-specific question, 
failing to leverage available household data & Count \\
Appropriate general response & Count of responses correctly providing a conceptual explanation for a general 
knowledge question & Count \\
Technical terms explained & Count of domain-specific terms (e.g., TOU, load shifting, kWh) defined or 
explained when first introduced & Count \\
Information-before-action rate & Proportion of control actions preceded by the system communicating
relevant device state or contextual information to the user before executing the action & \% (0--100\%) \\
Action confirmation rate & Proportion of control actions followed by explicit confirmation of the completed 
action & \% (0--100\%) \\
Action explanation rate & Proportion of control actions accompanied by an explanation of why the setting 
or schedule was chosen & \% (0--100\%) \\
\midrule
\multicolumn{3}{l}{\textbf{Control Capability}} \\
Target device accuracy & Proportion of device changes affecting only the intended target device(s) without 
unintended side effects & \% (0--100\%) \\
Schedule correctness & Proportion of load-shifting actions scheduled during off-peak hours per the utility's 
TOU rate structure. Energy-reducing actions during peak periods (e.g., load shedding) are counted as correct; 
demand response scenarios are excluded, as a DR response must occur within the peak event window & \% (0--100\%) \\
Mode correctness & Proportion of device mode changes appropriate for current conditions (e.g., cooling mode 
during summer) & \% (0--100\%) \\
Constraint compliance rate & Proportion of control actions respecting device-specific operational limits 
(e.g., valid temperature range) & \% (0--100\%) \\
\midrule
\multicolumn{3}{l}{\textbf{System Efficiency}} \\
Avg.\ system response length & Mean character length of all system responses within a conversation & Characters \\
Response latency & Mean time between receiving a user query and returning the system response, measured at 
the application level & Milliseconds \\
Token usage & Total input and output tokens consumed by the system during a conversation & Count \\
\end{longtable}
\end{small}
  
Personalization-related metrics under Interaction Quality evaluate whether the system tailors responses to match question
type. Data-specific questions should receive data-backed responses that leverage user-specific information; general
questions should receive conceptual explanations without unnecessary personalization (Table~\ref{tab:response_matrix}).
The off-diagonal cells capture mismatches: under-personalization occurs when the system provides generic answers
to data-specific questions, while over-personalization occurs when the system injects user data into responses to
general questions. This matrix measures whether the system recognizes when personalization adds value.

\begin{table}[b]
\centering
\caption{Response appropriateness matrix, classifying system responses by alignment between question type and response type}
\label{tab:response_matrix}
\small
\begin{tabular}{lcc}
\toprule
 & Data-specific question & General question \\
\midrule
Data-backed response & Appropriate data-backed response & Over-personalized response \\
General response & Under-personalized response & Appropriate general response \\
\bottomrule
\end{tabular}
\end{table}

The framework has three design properties. First, GQM traceability ensures every metric traces to an
explicit goal and no goal is left unmeasured. Second, modular structure allows selective application:
the first three categories apply to any conversational HEMS, control capability applies only when device
control is present, and system efficiency applies universally. Third, diagnostic granularity -- separating
over- and under-personalization, and decomposing action transparency into three independent behaviors --
enables targeted diagnosis of \textit{where} and \textit{how} a system falls short.

\section{Framework Validation}
\label{sec:validation}

\subsection{Experimental Design}
\label{sec:experimental_design}

\begin{figure}
\centering
\includegraphics[width=\columnwidth]{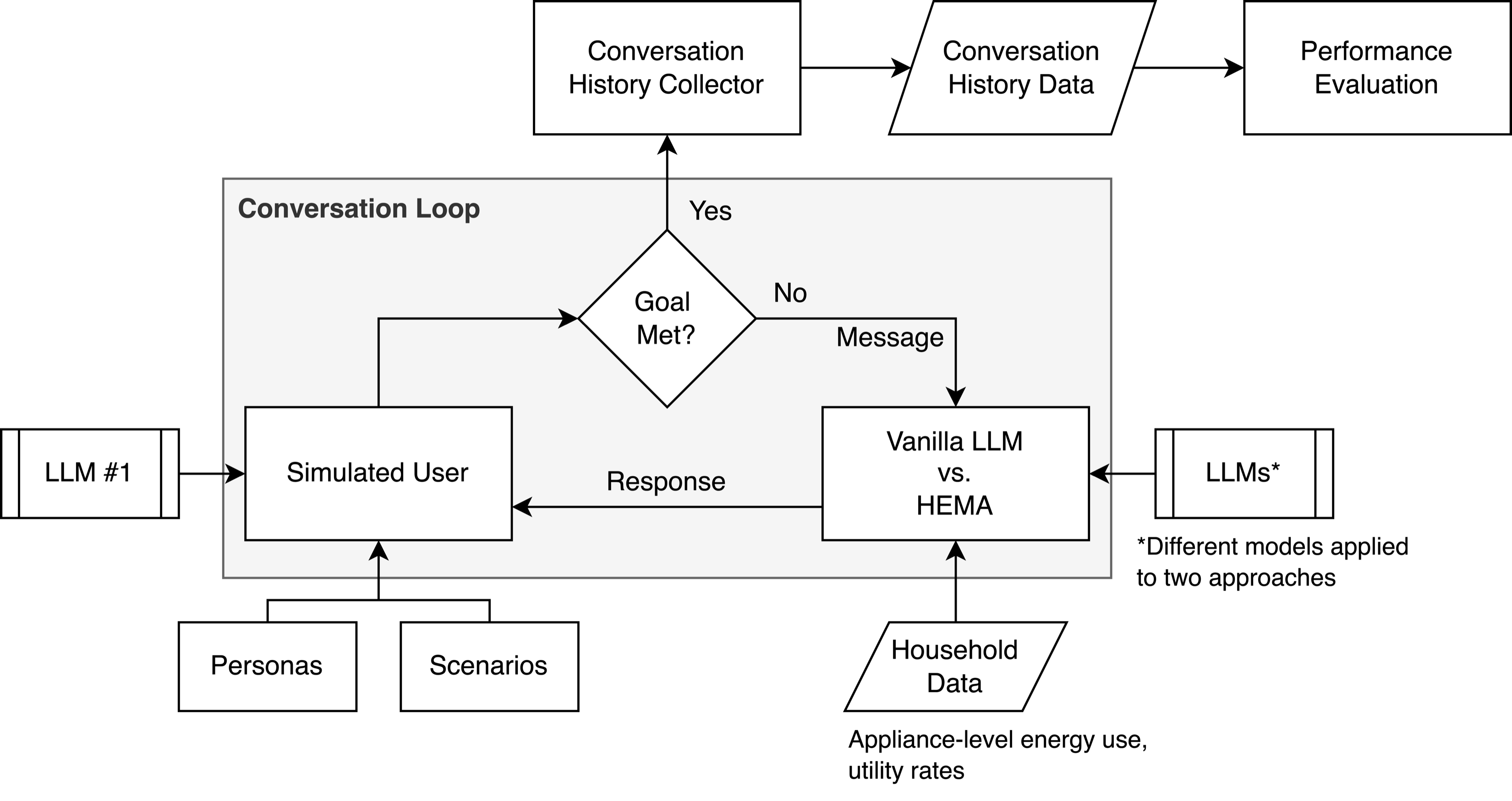}
\caption{The evaluation process of comparative analysis}
\label{fig:evaluation_process}
\end{figure}

To exercise the proposed 23-metric evaluation framework across a wide range of conversational conditions, a comparative 
analysis was conducted by generating and evaluating multi-turn conversations between simulated users and four system 
configurations, as shown in Fig.~\ref{fig:evaluation_process}.

\textbf{User}. An LLM-as-user approach was employed to generate multi-turn dialogues. The simulated user
role-played as a realistic persona using Gemini 2.0 Flash \cite{google2026gemini}, while the system under
test used OpenAI GPT-4o-mini \cite{openai2024gpt4omini}. Different LLMs were intentionally employed to
avoid implicit alignment risk; selections were informed by pilot tests for conversation flow coherence
and cost-effectiveness.

Each simulated user received a prompt containing persona traits and scenario objectives
(Tables~\ref{tab:personas}--\ref{tab:scenarios}). The nine analysis- and knowledge-focused scenarios
span energy data analysis, rate structure comprehension, and advisory tasks; the seven
control-focused scenarios cover three dimensions: control complexity, temporal scope, and
context sensitivity (e.g., off-peak operation or demand response event handling).

\begin{table}
\centering
\caption{Details of the personas considered in comparative analysis}
\label{tab:personas}
\small
\begin{tabular}{p{2.5cm}p{2cm}p{7cm}}
\toprule
Persona & Technical level & Description \\
\midrule
Confused newcomer & Novice & First-time homeowner unfamiliar with energy concepts \\
Tech-savvy optimizer & Expert & Software engineer interested in data-driven optimization \\
Budget-conscious parent & Intermediate & Single parent focused on reducing monthly expenses \\
Eco-conscious renter & Intermediate & Environmentally focused renter with limited control \\
Skeptical senior & Novice & Retired homeowner skeptical of technology \\
\bottomrule
\end{tabular}
\end{table}

\begin{table}
\centering
\caption{Details of the scenarios considered in comparative analysis}
\label{tab:scenarios}
\small
\begin{tabular}{p{3cm}p{4cm}p{5.5cm}}
\toprule
Scenario & Description & Primary goal \\
\midrule
\multicolumn{3}{l}{\textbf{Analysis-focused scenarios}} \\
Appliance analysis & Identify energy-consuming appliances & Find top energy consumers and get reduction recommendations \\
Peak reduction strategy & Reduce peak hour usage & Develop strategy to shift usage to off-peak hours \\
HVAC optimization & Optimize heating/cooling efficiency & Learn optimal settings, scheduling, and maintenance tips \\
Energy comparison & Compare usage over time & Compare energy usage between different periods \\
Solar consideration & Evaluate solar panel viability & Understand costs, savings, and factors affecting solar ROI \\
\midrule
\multicolumn{3}{l}{\textbf{General knowledge-focused scenarios}} \\
Understand utility rate & Learn TOU rate structure & Understand what TOU pricing means and when rates are highest \\
Weather energy impact & Understand weather effects on energy & Learn how weather affects bills and get weather-based advice \\
Utility rate details & Get specific rate information & Learn exact tier prices, peak hours, and seasonal differences \\
Rebate inquiry & Find rebate programs & Learn about available rebates, eligibility, and how to apply \\
\midrule
\multicolumn{3}{l}{\textbf{Control-focused scenarios}} \\
Thermostat adjustment & Adjust smart thermostat & Set optimal temperature based on rate structure \\
EV charging schedule & Schedule EV charging & Configure charging during off-peak hours \\
Device status check & Review all smart devices & Get overview of device status and energy consumption \\
Water heater optimization & Optimize water heater schedule & Align heating schedule with TOU rates while ensuring 
hot water availability \\
Pool pump scheduling & Schedule pool pump operation & Set pump runtime during optimal hours for savings \\
Vacation preparation & Configure multiple devices for vacation & Set up HVAC, water heater, and pool pump to minimize 
energy while away \\
Demand response event & Handle demand response notification & Adjust devices to reduce load during event period while 
maintaining comfort \\
\bottomrule
\end{tabular}
\end{table}

During each conversation, the simulated user generated short, natural responses of one to three sentences, 
exhibiting realistic behaviors throughout -- asking clarifying questions when confused, expressing 
frustration with unclear explanations, and maintaining persona-consistent communication patterns. The 
simulated user signaled goal completion when the key objective was satisfied, allowing conversation 
termination. Another conversation termination case was when the maximum turn limit (set at 40) was 
reached. This was treated as an unsuccessful interaction in which the system failed to guide the user 
toward goal completion within a reasonable conversation horizon.

The LLM-as-user methodology serves as a system diagnostic tool rather than a substitute for human user studies.
Prior work in this domain reports 90.5\% reproducibility and successful discrimination between system
configurations \cite{jung2026llmasuser}; broader HCI research confirms that LLM-simulated users can
approximate human behavioral distributions in structured task settings
\cite{yoon2024evaluating,sekulic2024reliable,hu2024quantifying}, supporting evaluation at the scale of
970 conversations that would be impractical with human participants alone.

\textbf{System Configuration}. Four HEMS configurations representing varying architectural sophistication were as follows:

\begin{enumerate}
    \item A vanilla LLM fed with the raw energy data (Vanilla+Raw).
    \item A vanilla LLM fed with a comprehensive energy data analysis report (Vanilla+Energy Report).
    \item A vanilla LLM fed with the comprehensive energy data analysis report and guided by a chain-of-thought
    (CoT) prompt (Vanilla+Energy Report+CoT).
    \item A multi-agent architecture of HEMS, named Home Energy Management Assistant (HEMA) \cite{jung2026cot}
    (more details of HEMA are provided in Appendix~\ref{app:hema}).
\end{enumerate}

These form a deliberate sophistication gradient: from unstructured data access (Vanilla+Raw), to structured but 
prompt-free access (Vanilla+Energy Report), to structured access with explicit CoT guidance (Vanilla+Energy Report+CoT), 
and finally to the multi-agent architecture. A framework with discriminative validity should detect performance 
differences proportional to this gradient across the relevant evaluation dimensions.

This comparative analysis utilized two weeks of data from a single household (a 302.4\,m\textsuperscript{2}
single-family house built in 2011 in Austin, Texas, with solar panels, an EV charger, a pool pump, an HVAC
unit, and 14 total appliances; data from Pecan Street Inc.\ \cite{pecanstreet2025}). The two-week duration
was constrained by the vanilla LLM's token limitations when processing raw time-series inputs. The use of a single household reflects this study's goal: evaluating 
the system architecture's ability to classify queries, coordinate specialized responses in tool outputs, and sustain 
multi-turn interaction -- capabilities that are properties of the AI system, not of the household data. This is 
consistent with design science methodology, in which a single artifact instantiation evaluated in its operational 
context constitutes a recognized evaluation approach \cite{hevner2004design}. Moreover, He and
Jazizadeh \cite{he2025context} empirically demonstrated that their LLM-based energy management system produced no
statistically significant performance differences across four residential buildings in different climate zones
(ANOVA $p > 0.05$). While that study evaluated a different system architecture, the finding suggests that
building-specific data may not substantially influence system-level evaluation outcomes.

The selected household was chosen for its feature richness, including solar generation, an EV charger,
a pool pump, and an HVAC unit, among 14 total appliances, to exercise all five evaluation categories across their full
scope. To enable control-related scenarios, a simulated smart home environment was developed that closely
replicated the structure and behavior of commercial smart home appliance APIs. 
The device inventory was derived from actual household appliance data. Device control capabilities and state 
properties were defined based on smart home platform integration specifications, with controllable features 
verified against publicly available developer documentation \cite{google2025home,apple2026homekit,samsung2026smartthings}. 
Each device, represented as a JSON object, specified five elements: device metadata (ID, connectivity, smart
control flag), supported capabilities (scheduling, variable-speed, demand response), current state (power,
mode, setpoints, sensor readings), configurable settings with valid ranges (e.g., thermostat 60--85$^\circ$F),
and control actions as action-parameter pairs analogous to REST API endpoint definitions.

To ensure a fair comparison, the two vanilla LLM configurations with structured data (the second and third
configurations) received a pre-computed energy analysis report generated using the analysis tools available to HEMA. This
ensured that these configurations had access to the same analytical results that HEMA would produce
during its tool-mediated workflow. The CoT prompt was designed to reflect HEMA's query classification logic and response guidelines.

\textbf{Dialogue Generation}. Each test combination was run five times to accommodate the LLM variability. For 
analytical and informative capabilities, a total of 225 test cases (= five personas $\times$ nine scenarios 
[five analysis-focused and four knowledge-focused] $\times$ five repetitions) were conducted per each system 
configuration. For control capability, HEMA was tested across 70 test cases (= two personas $\times$ seven 
control scenarios $\times$ five repetitions). The two personas were selected to represent the user technical 
proficiency spectrum: the tech-savvy optimizer (expert-level) and confused newcomer (novice-level) because 
the primary evaluation criterion for control-focused scenarios was whether the requested device action was 
executed correctly rather than how the system tailored its explanation. By selecting personas at opposite ends 
of the proficiency spectrum, this evaluation design served as a verification that the system's control capabilities
function consistently regardless of user characteristics, while still capturing any persona-driven differences
in the conversational interaction leading to the control action. HEMA's total run count was therefore 295.

When the conversation ended, the dialogue data were manually assessed to ensure their conversational flow -- 
whether it generated reasonable human-AI interactions. Conversations that exhibited degenerate patterns 
(e.g., repetitive loops, nonsensical user turns, or premature termination unrelated to goal completion) were 
discarded and regenerated. A total of 15 dialogues out of 970 required regeneration, and no conversation was 
regenerated more than once.

\subsection{System Performance Evaluation}
\label{sec:evaluation}
 
The system performance evaluation procedure followed a two-stage design. Of
the 23 metrics, eight were computed programmatically from system logs: turns
to completion, average system response length, response latency, and token
usage from conversation metadata, and target device accuracy, schedule
correctness, mode correctness, and constraint compliance rate from
deterministic comparison of device states before and after control actions.
The remaining 15 metrics required evaluative judgment rather than
direct computation from system logs; accordingly, an automated LLM-as-judge
pipeline was utilized to score all conversations at scale.
However, before cross-configuration results from this pipeline could be interpreted, its reliability had to first
be established. Stage~1 therefore conducted a human validation study on a stratified subset of conversations to
quantify agreement between human coders and automated scores. Stage~2 then applied the validated pipeline across
all four system configurations. This ordering ensures that substantive findings rest on empirically verified
measurement reliability.

\textbf{Human coder validation}. To validate the LLM-as-judge pipeline and identify which metrics are most
amenable to automation, 25 conversations spanning all scenario types and user personas were independently
rated by three trained coders, yielding 75 annotation sets. The sample provides comprehensive coverage
of the framework's measurement scope and is consistent in scale with comparable conversational AI
evaluation studies \cite{mehri2020usr,zheng2023judging,liu2023geval}.

Human validation was conducted on HEMA conversations specifically; since the identical rubrics and scoring protocol 
are applied across all system configurations, reliability demonstrated on HEMA conversations is assumed to generalize 
to the other configurations. One conversation was selected per sampled persona-scenario combination, choosing the run 
closest to the median Gemini 2.5 Flash score for that combination to ensure representativeness without cherry-picking. 
Coders were trained using the scoring method and practice conversations before independent rating, each requiring 
approximately 3--4 hours of independent annotation. Following initial independent scoring, an iterative calibration 
process was conducted: metric-level disagreements were reviewed, scoring rubrics were clarified where ambiguities 
were identified, and coders re-scored affected metrics under the refined rubrics. This calibration process is 
consistent with established content analysis methodology, where independent coding followed by structured
reconciliation improves measurement validity \cite{krippendorff2018content,campbell2013coding}. This process
produced adjudicated consensus scores for each metric. Agreement between the LLM evaluator and human consensus
was assessed using ICC(2,1) (two-way random, single measures; \cite{shrout1979intraclass}) for continuous metrics
and Cohen's kappa for the binary goal achievement metric. Mean absolute error (MAE) was computed as a
supplementary measure of agreement magnitude. Disagreement cases were analyzed to identify which evaluation
dimensions are harder to automate, yielding practical guidance for future applications of the framework.

\textbf{LLM-as-judge}. The human validation study served a dual purpose: establishing measurement
reliability and calibrating the automated pipeline. Where systematic LLM-human disagreements were
identified, scoring rubrics were revised and the pipeline was re-run on the 25 validation conversations
until satisfactory agreement was achieved. The calibrated pipeline -- using Gemini 2.5 Flash
(temperature\,=\,0) with structured extraction prompts, chain-of-thought reasoning, and explicit
quantitative calculations \cite{zheng2023judging} -- was then applied to score all 970 conversations
across the four system configurations. To stabilize residual stochastic variation in semantic extraction
(user questions, technical terms, answered question ratio), a three-run median voting procedure was
applied: each conversation was evaluated three times and the median count was selected for each extracted
category. Eight metrics were computed programmatically from system logs without LLM involvement: turns to
completion, average system response length, response latency, and token usage from conversation metadata,
and target device accuracy, schedule correctness, mode correctness, and constraint compliance rate from
deterministic comparison of device states before and after control actions.

Results are reported in two stages in the following section. Human validation results -- per-metric
LLM-versus-human consensus agreement -- are presented first to establish the reliability of the automated
pipeline before substantive findings are interpreted. Cross-configuration comparison results are then reported by
evaluation dimension (task performance, factual accuracy, interaction quality, control capability, and system
efficiency), with reliability levels from the human validation informing the confidence with which each set of
findings is discussed.

\section{Results}
\label{sec:results}

\subsection{Human Validation Results}
\label{sec:human_validation}

Three coders independently scored 25 HEMA conversations across the
framework's metrics, followed by a three-round calibration process to
resolve disagreements. Round~1 consisted of independent scoring.
Round~2 addressed the 11 conversations with a coder-range $\geq2$ on
personalization metrics, establishing seven scoring rules (e.g.,
question-type classification determines response-type classification;
rate and tariff information is general knowledge; counting is performed
at the turn level, not the sub-bullet level). Round~3 reviewed all 25
conversations under the refined rubrics; two scores were revised by
majority agreement, yielding unanimous consensus on all 25 conversations
for the four personalization metrics. For the remaining metrics, final
scores were determined by majority vote (categorical) or mean of three
coders (continuous).

One conversation was excluded due to an LLM extraction failure, yielding
24 conversations for the interaction quality analysis; factual accuracy
metrics were validated on the subset of conversations containing
verifiable numerical claims ($N$\,=\,12--25, depending on the metric).
Table~\ref{tab:validation} reports agreement between the LLM evaluator
and human consensus for all 15 LLM-scored metrics.

\begin{table}
\centering
\caption{LLM-as-judge validation: per-metric agreement with human consensus (post-calibration)}
\label{tab:validation}
\footnotesize
\renewcommand{\arraystretch}{1.1}
\begin{tabularx}{\textwidth}{p{4.2cm}p{1.2cm}Xp{2.0cm}Xp{1.8cm}}
\toprule
Metric & Type & $N$ & ICC(2,1) / $\kappa$ & MAE & Reliability \\
\midrule
goal\_achieved & Binary & 24 & $\kappa = 1.000$ & 0.000 & Perfect \\
under\_personalized & Count & 24 & 1.000 & 0.000 & Perfect \\
action\_explanation\_rate & Rate & 8 & 1.000 & 0.000 & Perfect \\
mean\_error\_pct & Rate & 12 & 0.970 & 0.070 & Excellent \\
factual\_accuracy\_rate & Rate & 12 & 0.935 & 0.012 & Excellent \\
appropriate\_general & Count & 24 & 0.907 & 0.250 & Excellent \\
user\_questions & Count & 24 & 0.897 & 0.542 & Good \\
factual\_claims & Count & 25 & 0.845 & 2.680 & Good \\
accurate\_claims & Count & 25 & 0.839 & 2.640 & Good \\
appropriate\_data\_backed & Count & 24 & 0.808 & 0.375 & Good \\
answered\_q\_ratio & Rate & 24 & 0.804 & 0.040 & Good \\
tech\_terms\_explained & Count & 24 & 0.734 & 0.500 & Moderate \\
\midrule
info\_before\_action\_rate & Rate & 8 & 0.464\textsuperscript{*} & 0.281 & -- \\
over\_personalized & Count & 24 & --\textsuperscript{*} & 0.083 & -- \\
action\_confirmation\_rate & Rate & 8 & --\textsuperscript{*} & 0.041 & -- \\
\bottomrule
\multicolumn{6}{p{0.95\textwidth}}{\footnotesize \textsuperscript{*}Restricted range: near-zero variance in human consensus scores renders ICC unreliable; MAE is the informative measure (see text). ICC\,=\,ICC(2,1), two-way random, single measures \cite{shrout1979intraclass}; $\kappa$\,=\,Cohen's kappa; MAE\,=\,mean absolute error. $N$ varies by metric: 25 for factual claim counts (all conversations), 24 for interaction quality metrics (one conversation excluded due to LLM extraction failure), 12 for factual accuracy rates (conversations with non-zero factual claims), 8 for control-specific metrics (one control conversation lacked device control actions).}
\end{tabularx}
\renewcommand{\arraystretch}{1.0}
\end{table}

\textbf{Metrics with sufficient variance.} Of the 15 LLM-scored metrics, twelve exhibited sufficient
variance for ICC to be computed: three achieved perfect agreement (ICC\,=\,1.000: under-personalized
response, action explanation rate, goal achievement); two factual accuracy metrics achieved excellent
agreement (ICC\,$\geq$\,0.935); four achieved good agreement (ICC\,=\,0.808--0.907); and technical
terms explained achieved moderate agreement (ICC\,=\,0.734). Full per-metric values are in
Table~\ref{tab:validation}.

\textbf{Restricted-range metrics.} Three metrics -- over-personalized response, action confirmation rate,
and information-before-action rate -- exhibited near-zero variance in human consensus scores, rendering
ICC mathematically inappropriate. Over-personalized response received a consensus score of zero across all
24 conversations; action confirmation rate received a consensus score of 1.0 in seven of eight control
conversations. Under these conditions, even a single minor disagreement produces an ICC at or below zero
despite trivially small absolute differences. This restricted-range limitation is well-documented in the
ICC literature \cite{koo2016guideline}. For these metrics, MAE provides a more informative measure of
agreement: 0.041 for action confirmation rate (near-ceiling human scores), 0.083 for over-personalized
response (near-floor human scores), and 0.281 for information-before-action rate.

\textbf{Factual accuracy metrics.} The four factual accuracy metrics were validated on conversations
with at least one verifiable numerical claim. High agreement on rate metrics (ICC $\geq$ 0.935) indicates
that the LLM pipeline identifies similar numbers of claims as human coders and agrees on which are
accurate.

\textbf{Voting stabilization.} The three-run median voting procedure described in Section~\ref{sec:evaluation}
improved agreement for metrics exhibiting stochastic variation: answered question ratio improved from
moderate (ICC\,=\,0.593 without voting) to good agreement (ICC\,=\,0.804), and technical terms explained
stabilized from ICC values fluctuating between 0.3 and 0.95 across runs to a stable 0.734.

In summary, twelve of fifteen LLM-scored metrics achieved ICC(2,1) $\geq$ 0.73 (moderate to excellent),
with three achieving perfect agreement and two achieving excellent agreement above 0.93 on factual
accuracy. For the remaining three metrics with restricted range in human scores, MAE confirmed close
agreement (0.041 to 0.281). These results support the use of the automated pipeline for scoring all 970
conversations across the four system configurations, with the caveat that over-personalized response and
action confirmation rate -- both near-constant across the validation sample -- should be interpreted with
awareness of the limited variance available for validation. The eight programmatic metrics (four system
efficiency and four control capability) are computed deterministically from system logs and do not require
LLM-based validation. The validated pipeline was then applied to score all 970 conversations, the results
of which are reported in the following subsections.

\subsection{Analytical and Informative Capabilities}
\label{sec:analytical}

\begin{figure}
\centering
\includegraphics[width=\textwidth]{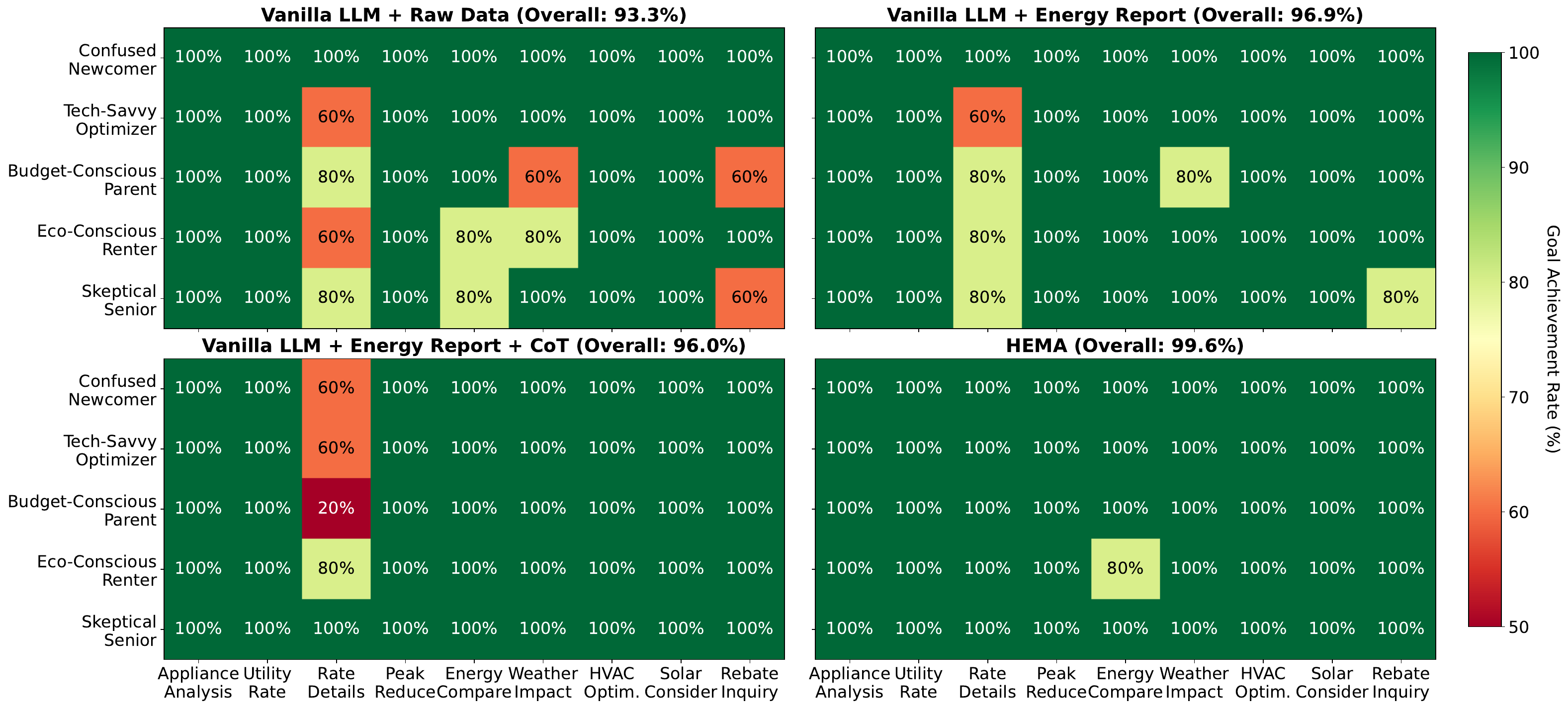}
\caption{Goal achievement rates for each system configuration across five user personas and nine scenarios}
\label{fig:goal_achievement}
\end{figure}

\textbf{Task Performance}: Figure~\ref{fig:goal_achievement} illustrates goal achievement rates across four system configurations, five simulated user personas, and nine evaluation scenarios. All configurations achieved high overall goal achievement rates, ranging from 93.3\% (Vanilla+Raw) to 99.6\% (HEMA), with Vanilla+Energy Report at 96.9\% and Vanilla+Energy Report+CoT at 96.0\%. A chi-squared test confirmed significant differences across configurations ($\chi^2 = 12.96$, $p = 0.005$), with post-hoc Fisher's exact tests indicating that the only significant pairwise difference after Bonferroni correction was between Vanilla+Raw and HEMA ($p = 0.003$). The heatmaps reveal that these aggregate differences are driven by scenario-specific failures.

All systems performed consistently well on five of nine scenarios -- appliance analysis, HVAC optimization, peak reduction strategy, solar consideration, and understand utility rate -- achieving 100\% goal achievement across all configurations. However, the utility rate details scenario exposed a critical limitation of the baseline approaches: without access to domain-specific knowledge, baseline systems could not provide specific utility rate tier information when requested. When users requested Austin Energy's tiered pricing structure, baseline systems responded with ``I do not have the specific rates'' and directed users to external resources, causing premature conversation termination. Baseline goal achievement on this scenario ranged from 64\% (Vanilla+Energy Report+CoT) to 80\% (Vanilla+Energy Report), while HEMA achieved 100\% by retrieving rate structure data using its Knowledge agent.

HEMA demonstrated near-perfect consistency across persona-scenario combinations, with only a single cell below 100\%: eco-conscious renter on energy comparison (80\%). In contrast, Vanilla+Energy Report+CoT exhibited one cell as low as 20\% (budget-conscious parent on utility rate details), and Vanilla+Energy Report ranged from 60\% to 100\%.

Regarding turns to completion, HEMA used a comparable number of conversational turns ($6.62 \pm 2.45$) to baseline systems ($5.74 \pm 1.94$ to $6.35 \pm 2.24$) while achieving higher goal completion rates. The differences were statistically significant ($H = 12.38$, $p = 0.006$) but small in magnitude, indicating that HEMA's conversations were productive interactions that successfully guided users toward their HEM goals within a similar conversational footprint.

\textbf{Factual Accuracy}: Table~\ref{tab:factual_accuracy} presents the factual accuracy metrics across all four system configurations. A striking distinction emerged between the Vanilla LLM+Raw approach and all structured data configurations. The Vanilla+Raw system achieved only 56.4\% factual accuracy with an exceptionally high mean error of 30.9\%, indicating that when the LLM attempts to interpret raw energy data directly, it frequently produces incorrect or fabricated numerical claims. Analysis of individual claims revealed several patterns of failure:

\begin{itemize}
    \item \textbf{Cost fabrication}: The system claimed ``\$100/month'' savings when actual potential was \$12.84 (679\% error).
    \item \textbf{Appliance ranking errors}: Pool pumps were consistently under-ranked (5th--6th when actually 2nd largest consumer).
    \item \textbf{Magnitude errors}: Daily usage reported as ``30\,kWh'' when actual consumption was 138.6\,kWh (78\% error).
\end{itemize}

In contrast, all systems utilizing structured energy analysis reports or analysis tools achieved factual accuracy rates at or above 95.8\%, with mean errors below 1\%.

\begin{table}
\centering
\caption{Factual accuracy metrics across system configurations}
\label{tab:factual_accuracy}
\footnotesize
\begin{tabularx}{\textwidth}{lXXXX}
\toprule
Metric & V+Raw & V+Report & V+Report+CoT & HEMA \\
\midrule
Factual claims & $0.82 \pm 1.73$ & $4.23 \pm 5.30$ & $4.73 \pm 6.22$ & $7.19 \pm 9.05$ \\
Accurate claims & $0.42 \pm 0.96$ & $4.14 \pm 5.24$ & $4.67 \pm 6.17$ & $7.02 \pm 8.99$ \\
Factual accuracy rate\textsuperscript{†} & $56.4\% \pm 35.8\%$ & $98.0\% \pm 8.5\%$ & $98.4\% \pm 5.6\%$ & $95.8\% \pm 9.7\%$ \\
Mean error (\%)\textsuperscript{†} & $30.9 \pm 53.0$ & $0.4 \pm 2.4$ & $0.8 \pm 4.2$ & $0.8 \pm 2.3$ \\
\bottomrule
\multicolumn{5}{l}{Values show mean $\pm$ standard deviation.} \\
\multicolumn{5}{l}{\textsuperscript{†}Computed only on conversations with $\geq$1 verifiable numerical claim ($N$ = 58/112/108/113 per configuration).} \\
\multicolumn{5}{l}{V+Raw = Vanilla+Raw; V+Report = Vanilla+Energy Report; V+Report+CoT = Vanilla+Energy Report+CoT.} \\
\end{tabularx}
\end{table}

Notably, we observed minimal variation in factual accuracy rates across different user personas and conversation 
scenarios. This consistency suggests that factual accuracy was primarily determined by the system architecture 
(specifically, whether structured analysis reports are used) rather than by conversational context.

HEMA generated 7.19 factual claims per conversation versus 4.73 for Vanilla+Energy Report+CoT and 4.23
for Vanilla+Energy Report -- a 52\% increase in data-driven content -- while maintaining comparable
accuracy (7.02 accurate claims), indicating that its multi-agent architecture enables richer responses
without sacrificing reliability.

\begin{figure}
\centering
\includegraphics[width=\textwidth]{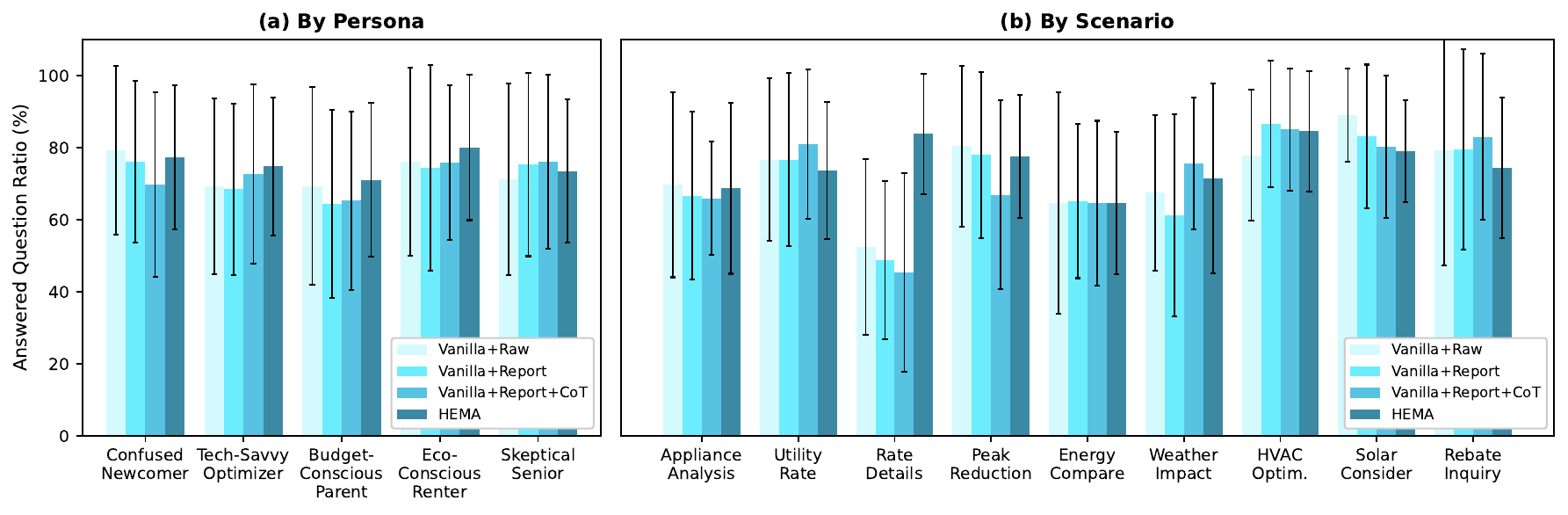}
\caption{Four system configurations' answered question rates by (a) persona and (b) scenario}
\label{fig:question_answer}
\end{figure}

\textbf{Interaction Quality}: The framework's interaction quality metrics revealed a two-level pattern. The
answered question ratio (Figure~\ref{fig:question_answer}) was comparable across all configurations -- ranging
from $71.7\% \pm 25.5\%$ (Vanilla+Energy Report) to $75.3\% \pm 20.2\%$ (HEMA) -- with no statistically
significant differences ($H = 2.21$, $p = 0.530$). This indicates that surface-level responsiveness was
maintained regardless of architectural sophistication. HEMA exhibited the lowest standard deviation, indicating
more consistent performance across diverse user types and consultation topics. The more diagnostically informative
differences emerge in the personalization calibration metrics below.

\begin{figure}
\centering
\includegraphics[width=\textwidth]{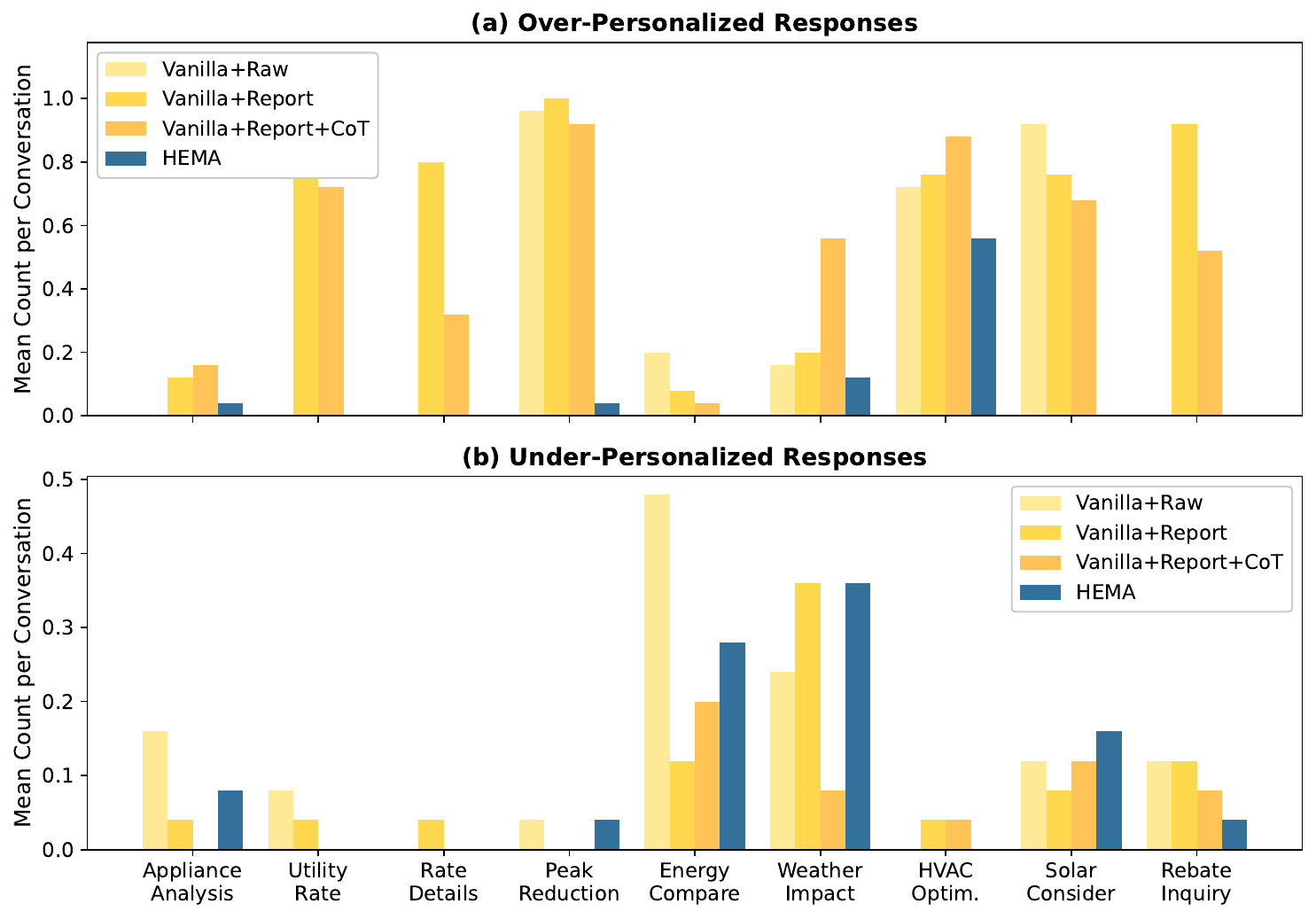}
\caption{Response appropriateness counts per conversation by scenario: (a) over-personalized responses and (b) under-personalized responses. Vanilla+Energy Report exhibits the highest over-personalization, concentrated in conceptual scenarios. HEMA achieves near-zero over-personalization across all scenarios.}
\label{fig:response_appropriateness}
\end{figure}

The response appropriateness metrics (Table~\ref{tab:response_appropriateness}; Figure~\ref{fig:response_appropriateness})
indicated distinct behavioral patterns across system configurations. Over-personalization showed the most
statistically significant variation ($H = 133.34$, $p < 0.001$). HEMA achieved the lowest over-personalization
($0.08 \pm 0.29$), indicating superior judgment about when data citation was warranted versus when general
explanations sufficed. This discriminative ability was also reflected in HEMA's highest appropriate general
response count ($1.69 \pm 1.20$, $H = 55.81$, $p < 0.001$); HEMA provided educational content rather than
unnecessarily referencing specific consumption data.

Vanilla+Energy Report exhibited the highest over-personalization ($0.64 \pm 0.65$), followed by
Vanilla+Energy Report+CoT ($0.53 \pm 0.60$) and Vanilla+Raw ($0.33 \pm 0.60$). This pattern indicates that
the structured energy report encouraged unnecessary injection of user-specific data into responses where a
general explanation would have been more appropriate -- a failure mode that HEMA's query routing mitigated.
Scenario-level analysis revealed that Vanilla+Energy Report's over-personalization concentrated in
conceptual scenarios: understand utility rate ($1.12$ versus HEMA: $0.00$), peak reduction strategy
($1.00$ versus HEMA: $0.04$), and rebate inquiry ($0.92$ versus HEMA: $0.00$).

Under-personalization was low across all configurations ($0.06$--$0.14$) with no significant differences
($H = 5.32$, $p = 0.150$), suggesting that all systems adequately leveraged available data when users
inquired about specific energy patterns.

\begin{table}
\centering
\caption{Response appropriateness results across system configurations}
\label{tab:response_appropriateness}
\footnotesize
\begin{tabularx}{\textwidth}{lXXXX}
\toprule
Metric & V+Raw & V+Report & V+Report+CoT & HEMA \\
\midrule
Approp.\ data-backed resp. & $0.70 \pm 0.76$ & $0.68 \pm 0.80$ & $0.75 \pm 0.80$ & $0.73 \pm 0.89$ \\
Over-personalized resp. & $0.33 \pm 0.60$ & $0.64 \pm 0.65$ & $0.53 \pm 0.60$ & $0.08 \pm 0.29$ \\
Under-personalized resp. & $0.14 \pm 0.44$ & $0.09 \pm 0.37$ & $0.06 \pm 0.25$ & $0.11 \pm 0.37$ \\
Approp.\ general resp. & $1.42 \pm 1.32$ & $0.93 \pm 0.89$ & $1.07 \pm 1.05$ & $1.69 \pm 1.20$ \\
\bottomrule
\multicolumn{5}{l}{Values show mean $\pm$ standard deviation. Abbreviations as in Table~\ref{tab:factual_accuracy}.} \\
\end{tabularx}
\end{table}

Cross-system analysis revealed consistent persona patterns: tech-savvy optimizer received the most data-backed
responses across all configurations, while confused newcomer and budget-conscious parent received proportionally
more general explanations. Data-intensive scenarios (appliance analysis, peak reduction) generated 3--5 times
more data-backed responses than general inquiry scenarios (rebate inquiry, solar consideration). Notably,
Vanilla+Energy Report's over-personalization concentrated precisely in utility-rate scenarios, where users
asking conceptual questions received unnecessary data citations regardless of their technical level
(understand utility rate: $1.12$ per conversation versus HEMA: $0.00$).

\textbf{System Efficiency}: Table~\ref{tab:system_efficiency} presents the system efficiency results.

\begin{table}
\centering
\caption{System efficiency results derived from analysis and knowledge scenarios across four system configurations}
\label{tab:system_efficiency}
\footnotesize
\begin{tabularx}{\textwidth}{lXXXX}
\toprule
Metric & V+Raw & V+Report & V+Report+CoT & HEMA \\
\midrule
Mean response length & $1{,}810 \pm 549$ & $1{,}664 \pm 480$ & $1{,}381 \pm 399$ & $1{,}407 \pm 364$ \\
Mean response latency & $14.01 \pm 14.66$ & $10.08 \pm 23.18$ & $7.15 \pm 4.78$ & $14.10 \pm 3.48$ \\
Total tokens & $258{,}412 \pm 118{,}491$ & $4{,}454 \pm 2{,}417$ & $6{,}736 \pm 3{,}159$ & $37{,}703 \pm 17{,}011$ \\
\bottomrule
\multicolumn{5}{l}{Values show mean $\pm$ standard deviation. Abbreviations as in Table~\ref{tab:factual_accuracy}.} \\
\end{tabularx}
\end{table}

The system efficiency metrics revealed distinct trade-offs between the four configurations in terms of 
computational resources and response characteristics. Vanilla+Raw produced the longest responses, while 
HEMA and Vanilla+Energy Report+CoT generated more concise outputs. The lower standard deviation for HEMA 
also indicates more consistent response lengths across different scenarios and personas.

HEMA exhibited an average response latency of 14.10 seconds with notably low variance ($\pm$3.48 seconds), 
demonstrating predictable performance regardless of query complexity. In contrast, the baseline configurations 
had high latency variance -- particularly Vanilla+Energy Report -- indicating inconsistent processing times. 
The Vanilla+Energy Report+CoT system achieved the fastest average response time since it operated with a 
simpler single-pass architecture with detailed instructions.

The most striking difference appears in token usage. Vanilla+Raw consumed approximately 258,000 tokens per 
conversation since the entire raw energy dataset was included in each LLM context. The pre-processed Energy 
Report configurations dramatically reduced this to 4,500--6,700 tokens by summarizing the data before LLM 
interaction. HEMA uses approximately 37,700 tokens per conversation -- higher than the two baseline configurations
with the energy report, but far more efficient than raw data processing. This intermediate token usage reflects 
HEMA's agentic architecture, where multiple specialized agents and tool calls accumulated tokens across the 
conversation.

\subsection{Smart Device Control Analysis}
\label{sec:control}

Control capability metrics and action transparency metrics were computed exclusively for HEMA's 70 control 
test cases; vanilla LLM baselines were excluded as they do not support device control. Two control evaluation 
runs were conducted over the same 70 test cases. Time-of-use-dependent metrics -- schedule correctness and 
demand response handling -- are reported from the run whose simulated clock was aligned to the household energy 
dataset, so that rate periods matched the season represented in the data. The remaining metrics, which do not 
depend on rate periods, are reported from the run scored by the validated LLM-as-judge pipeline described in 
Section~\ref{sec:evaluation}. Both runs used identical personas, scenarios, and system configuration.

\textbf{Task Performance}: The framework's task performance revealed a 97.1\% goal success rate across seven 
control-focused scenarios, with failures concentrated in a single scenario type. In the device status check 
scenario, two cases arose where the Control agent provided a comprehensive device status table, but lacked 
quantitative power consumption values so did not fully meet the request.

\textbf{Action Correctness}: The target device accuracy metric revealed a high but imperfect $94.2\% \pm 12.7\%$ 
across 57 device control runs. All scenarios achieved 100\% except for vacation preparation (66.7\%), where 
the agent occasionally modified the pool pump in addition to the requested HVAC and water heater adjustments. 
This behavior reflects proactive energy optimization rather than task failure -- adjusting pool pump schedules 
during extended absence was a reasonable energy-saving measure, although it was not explicitly requested.

\begin{figure}
\centering
\includegraphics[width=0.65\columnwidth]{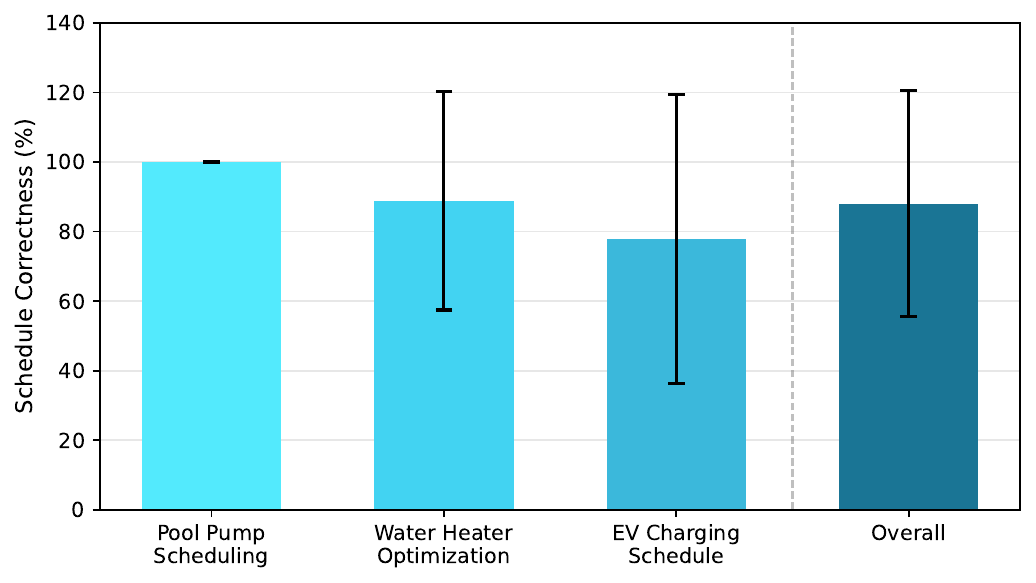}
\caption{Schedule correctness by load-shifting control scenario. Demand response is excluded, as a DR response must occur within the peak event window (see text).}
\label{fig:schedule_correctness}
\end{figure}

Schedule correctness was computed over the three load-shifting scenarios -- EV charging, pool pump scheduling, 
and water heater optimization -- for which off-peak placement is the correct objective. Across 25 runs, HEMA 
achieved $88.0\% \pm 32.5\%$ overall (Figure~\ref{fig:schedule_correctness}). Pool pump scheduling achieved 
100\%. Water heater optimization achieved $88.9\% \pm 31.4\%$: in one case the agent scheduled the water heater 
to start at 17:00, inside the afternoon peak window, when off-peak rates were available overnight. EV charging 
achieved $77.8\% \pm 41.6\%$; the two failures involved ``stop charging'' actions scheduled at 14:00 and 17:00. 
While halting charging during peak is defensible as cost avoidance, the scheduled times fell inside the peak 
window rather than completing the charge off-peak.

Demand response events are reported separately, because schedule correctness does not apply to them: a DR 
response must occur within the peak event window by definition, so off-peak placement would indicate failure 
rather than success. The appropriate criterion is whether the system completes the full event cycle. Across 
10 DR runs, HEMA applied a setback within the event window and scheduled a return to normal operation 
afterwards in nine cases (90\%). The single exception applied a shallower setback (76$^\circ$F rather than 
the 78$^\circ$F used in the other runs) and scheduled no return to normal, leaving the household in a setback 
state after the event ended.

Mode correctness and constraint compliance both achieved 100\%, indicating the agent consistently selected 
seasonally appropriate device modes (e.g., cooling in summer) and maintained all settings within safe operational 
boundaries (Figure~\ref{fig:schedule_correctness}). These ceiling effects suggest the agent reliably interpreted 
contextual factors and respects device constraints.

\begin{figure}
\centering
\includegraphics[width=0.75\columnwidth]{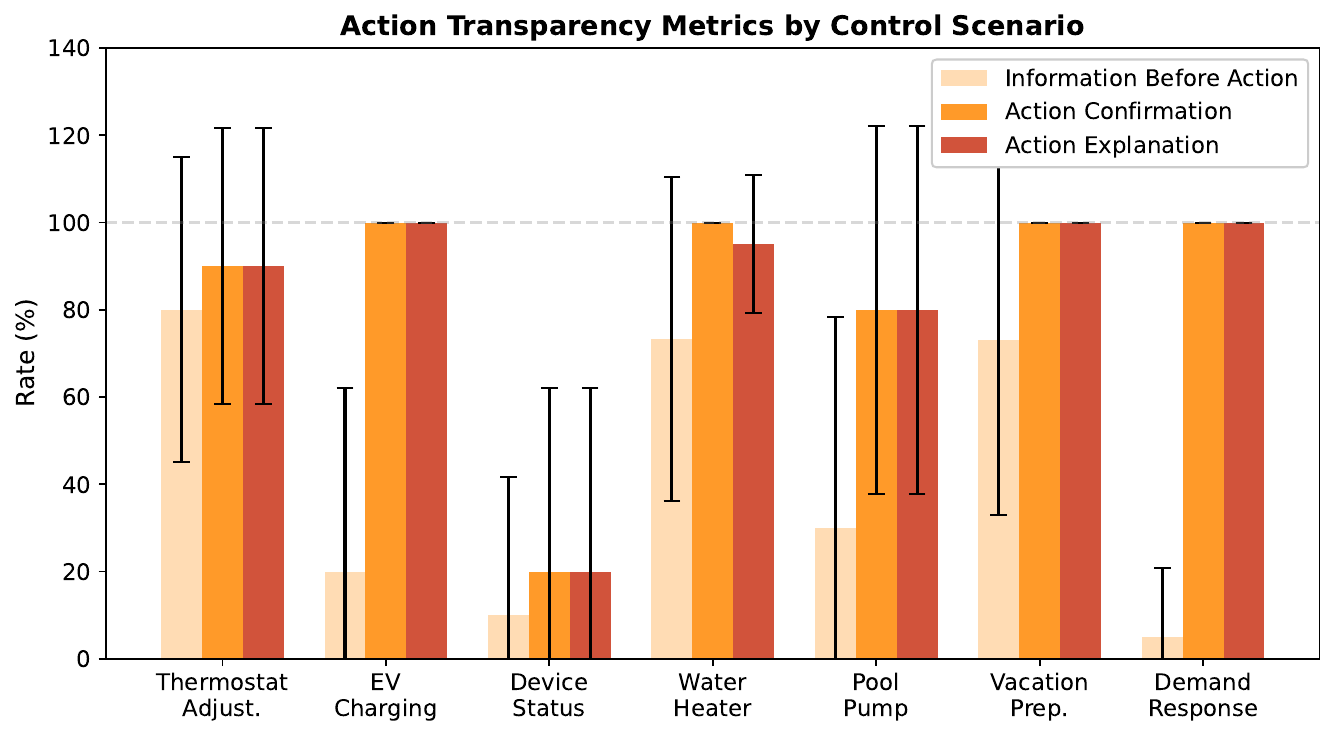}
\caption{Action transparency metrics (information-before-action rate, action confirmation rate, action explanation rate) by control scenario}
\label{fig:interaction_quality}
\end{figure}

\textbf{Interaction Quality}. Figure~\ref{fig:interaction_quality} presents the three action transparency
metrics across all seven control scenarios ($n = 70$). Action confirmation rate was the highest of the three
($84.3\% \pm 36.7\%$ overall), followed closely by action explanation rate ($83.6\% \pm 36.8\%$), while
information-before-action rate was substantially lower ($41.6\% \pm 46.6\%$). This pattern indicates that
the Control agent consistently communicated completed actions and their rationale to users, but frequently
proceeded to execute device commands without first gathering or communicating relevant device state information.

Scenario-level variation reveals distinct patterns by scenario type. Information-before-action was highest
for thermostat adjustment ($80.0\% \pm 35.0\%$) and vacation preparation ($73.0\% \pm 39.9\%$), where the
agent consistently retrieved device status before making changes. However, information-before-action dropped
sharply for demand response ($5.0\% \pm 15.8\%$) and device status check ($10.0\% \pm 31.6\%$), indicating
that in time-critical or status-only scenarios the agent acted without pre-action information gathering.
Action confirmation and action explanation were both at or near 100\% for demand response, EV charging,
vacation preparation, and water heater optimization. The notable exception was device status check, where
both confirmation ($20.0\% \pm 42.2\%$) and explanation ($20.0\% \pm 42.2\%$) were low -- consistent with
the informational nature of this scenario, where the primary action is status retrieval rather than device
control. This is practically relevant: users who do not understand why a charging or scheduling action was
set at a particular time may override it, undermining the energy-saving intent of the control action.

\textbf{System Efficiency}. Response length ranged from $483 \pm 62$ characters in the DR scenario to
$983 \pm 375$ in device status check, where comprehensive multi-device reports required longer outputs.
Response latency averaged 12.1 seconds across all scenarios. Token consumption ranged from
$17{,}687 \pm 1{,}660$ for DR -- a focused, single-action task requiring no prior information gathering --
to $121{,}672 \pm 102{,}974$ for device status check, where users triggered extensive energy analysis tool
calls across multiple devices. Scenarios requiring cross-domain reasoning (water heater optimization:
$83{,}163$ tokens; pool pump scheduling: $74{,}684$ tokens) consumed more tokens than single-domain actions
due to the agent's need to synthesize energy data, TOU rate context, and device state before acting.

\section{Discussion and Limitations}
\label{sec:discussion}

\subsection*{Necessity of Multi-Dimensional Evaluation of Conversational HEMS (RQ\#1)}

The results demonstrate that a single-metric or accuracy-only approach would provide an
incomplete and, in some cases, misleading characterization of conversational HEMS quality.
The 23-metric framework's discriminative validity is most directly evidenced by where
configurations that performed similarly on narrow metrics diverged substantially on broader
ones. Among the 23 metrics, factual accuracy effectively separated the Vanilla+Raw
configuration from all structured-data configurations but did not differentiate HEMA from
the report-based configurations, all of which exceeded 96\%. The more discriminating
dimension proved to be personalization calibration: Vanilla+Energy Report exhibited
$0.64 \pm 0.65$ over-personalized responses per conversation despite having access to
correctly structured, pre-computed analysis data, while HEMA achieved near-zero
over-personalization ($0.08 \pm 0.29$). This failure mode -- inappropriately injecting
user-specific data into responses to general knowledge questions -- persisted regardless
of data quality, revealing that architectural routing decisions, not data access, determine
whether general knowledge questions receive inappropriately data-specific responses. A
framework limited to task success and factual accuracy would classify Vanilla+Energy Report
and HEMA as equivalent performers while missing this qualitatively important behavioral
difference. Conversely, the answered question ratio was statistically flat across all four
configurations (71.7--75.3\%, $p = 0.530$) -- a null result that is itself informative:
surface-level question coverage was maintained regardless of architectural complexity,
indicating that this metric alone would fail to surface the substantial personalization
calibration differences the framework detects.

Multi-turn evaluation further strengthened the framework's diagnostic value by exposing
dynamics structurally inaccessible to single-turn benchmarks: goal achievement as a
sustained process rather than a one-shot output; per-turn productivity (HEMA achieved
higher completion rates within a comparable conversational footprint of 6.62 turns); and
persona-scenario interaction effects (e.g., the confused newcomer paired with the rate
details scenario as a uniquely challenging combination). Notably, Michelon et al.\
\cite{michelon2025llm} independently adopted a similar LLM-based user simulation approach
for evaluating their HEMS interface, suggesting emerging methodological acceptance in the
field.

The control capability category represents a third evaluation dimension that conventional
benchmarks omit entirely. The framework's four control metrics captured failure modes --
schedule misalignment, unintended device targeting -- that do not manifest in analytical
and informative test cases. Mode correctness and constraint compliance both achieved
100\%, establishing a safety floor, while schedule correctness ($88.0\% \pm 32.5\%$ across
load-shifting scenarios) and target device accuracy ($94.2\% \pm 12.7\%$) exposed residual
vulnerabilities even in a well-grounded architecture. Demand response required a separate
criterion altogether -- off-peak placement is a failure signal there, not a success one --
illustrating that control evaluation cannot be reduced to a single scheduling rule. These failure classes are architecturally
inaccessible to advisory-only systems and cannot be detected by response-quality metrics
alone.

\subsection*{Scalability of the LLM-as-Judge Approach (RQ\#2)}

The LLM-as-user approach enabled 970 conversations with controlled variation -- a scale
infeasible with human subjects. The use of different LLMs mitigated implicit alignment
risk, and structured persona traits with goal-completion criteria addressed known
reliability concerns \cite{sekulic2024reliable}.

The human validation study revealed that LLM-as-judge reliability is metric-dependent: twelve of fifteen
metrics achieved ICC(2,1) $\geq$ 0.73, with three restricted-range metrics confirmed via low MAE.
A key finding was that personalization metrics initially showed systematic LLM inflation due to a
counting-granularity mismatch, resolved through iterative rubric calibration. Researchers can confidently
automate most metrics at scale, while those involving subjective classification benefit from human
calibration. The iterative calibration protocol -- independent coding, structured reconciliation, and
rubric refinement -- is replicable for validating automated pipelines in domain-specific conversational AI.

\subsection*{Distinguishing Conversational HEMS Performance (RQ\#3)}

The primary strength of the agentic architecture was robustness to query diversity: HEMA
maintained 80\%+ goal achievement across nearly all persona-scenario combinations, while
non-agentic configurations showed high variance from 0\% to 100\%. This consistency stems
from the agentic system's ability to adapt its processing path per query, whereas the other
configurations applied a fixed strategy regardless of query type, making them vulnerable to
out-of-scope requests such as specific utility rate details, where all three configurations
failed while HEMA achieved 100\%. The agentic architecture also eliminated numerical
fabrication -- a well documented risk of LLM-generated content
\cite{huang2025hallucination} -- by grounding claims in tool outputs
\cite{schick2023toolformer} and demonstrated superior personalization judgment through
architectural separation between Analysis and Knowledge agents.

These strengths come with trade-offs. HEMA consumed approximately 37,700 tokens per
conversation -- 5.6 to 8.4 times more than report-based configurations -- and exhibited
higher response latency than the simpler Vanilla+Energy Report+CoT configuration. HEMA's
factual accuracy rate (95.8\%) was also marginally below Vanilla+Energy Report+CoT
(98.4\%), but this reflects a denominator effect: HEMA generated 52\% more factual claims
per conversation (7.19 vs.\ 4.73), and its absolute accurate claim count (7.02) exceeded
the next best configuration (4.67) by 50\% -- a higher volume of data-driven responses
proportionally increases exposure to minor numerical imprecision while delivering
substantially more verified content per conversation. The from-scratch development effort
across multiple system layers also represents a substantial barrier. Notably, for well-scoped, single-domain tasks (e.g., appliance analysis, general
concepts), non-agentic configurations performed comparably at lower cost. The agentic
architecture's value emerges for complex, multi-faceted, or knowledge-requiring
interactions, precisely the scenarios characterizing real-world sustained use.

Three architectural components drove HEMA's performance gains: (1)~the self-consistency
classifier \cite{wang2022selfconsistency} provided effective query routing -- its 99.0\%
goal achievement rate indicates that misrouting was rare, while non-agentic configurations'
single-context processing produced higher variance and failure on out-of-scope requests;
(2)~agent specialization calibrated personalization decisions more effectively than prompt
engineering alone, as demonstrated by the over-personalization contrast reported above;
and (3)~the ReAct paradigm \cite{yao2023react} grounded numerical claims in computed tool
outputs, eliminating fabrication. However, independent agent operation led to redundant tool
calls in cross-domain scenarios; future architectures should incorporate inter-agent
communication \cite{guo2024large}. Interaction quality metrics -- particularly
over-personalization -- proved more discriminating than factual accuracy for distinguishing
agentic from non-agentic architectures.

\subsection*{Practical Considerations for Deployment}

Remote LLM API access introduced latency and intermittent errors, motivating future local
or edge-deployed LLMs \cite{zheng2025edge}. Token constraints restricted the comparative
analysis to two weeks of data, particularly for Vanilla+Raw; HEMA's tool-mediated
architecture partially mitigates this by querying data on demand. LLM selection directly
affects latency, cost, and reasoning capability \cite{kaplan2020scaling}: GPT-4o-mini
yielded 14.10\,s average latency versus 23\,s for a GPT-4o-based HEMS \cite{he2025context},
though comparisons across architectures should be interpreted with caution.

\subsection*{Limitations}

Several limitations should be considered. First, the evaluation relied on simulated rather than real
occupants \cite{yoon2024evaluating,sekulic2024reliable}. LLM-simulated users may produce more cooperative
and less ambiguous queries than real users \cite{seshadri2026lost}, representing an optimistic upper bound
on system performance. To mitigate this, the five personas and 16 scenarios were informed by an empirical
study of real occupant interactions with a conversational HEMS \cite{jung2026humanai}, in which 156 prompts
from 20 participants were classified into six functional categories -- information seeking, goal articulation,
constraint communication, and solution evaluation -- ensuring the simulated personas cover behaviorally
grounded interaction patterns. Importantly, the 23-metric framework and its measurement procedures are not
dependent on simulated users and are directly applicable to real-user conversational data; future work should
replicate this evaluation with residential participants to verify that system rankings are stable across real
and simulated populations. The high aggregate goal achievement rates (93.3--99.6\%) may also partly reflect
the cooperative nature of simulated users; discriminating cases such as the utility rate details scenario
(64--80\% for baselines) and schedule correctness failures (77.8\% for EV charging) confirm that the
scenario set includes non-trivial cases where architectural differences remain clearly detectable.
Second, the evaluation draws on data from a single household in Austin, Texas \cite{pecanstreet2025}, chosen
to provide a controlled context in which ground-truth values can be precisely defined. The aim is to validate
the framework's metric operability and discriminant validity, not to generalize performance across the broader
U.S.\ household population; replication across diverse utility structures, climates, and device configurations
is a priority. Third, agents operated independently without inter-agent collaboration, leading to redundant
computation in cross-domain scenarios -- a design choice enabling independent attribution of errors to specific
agents. Fourth, evaluation used isolated sessions; longitudinal evaluation is needed to assess sustained
collaboration as user preferences evolve. Finally, the framework does not address actual energy savings,
long-term user satisfaction, or the optimality of underlying scheduling logic -- all acknowledged as
important directions for future evaluation extensions.

\section{Conclusion}
\label{sec:conclusion}

This study presented a comprehensive evaluation framework for conversational
HEMS, comprising 23 metrics organized across five categories -- task
performance, factual accuracy, interaction quality, control capability, and
system efficiency -- grounded in the Goal-Question-Metric (GQM) methodology.
The framework was validated through a comparative evaluation of four system
configurations representing a sophistication gradient, using 970 multi-turn
conversations generated via an LLM-as-simulated-user methodology across five
user personas and 16 scenarios.

The contributions are threefold. (RQ1)~A 23-metric GQM-grounded framework was proposed spanning task
performance, factual accuracy, interaction quality, control capability, and system efficiency, with
diagnostic granularity that distinguishes specific failure modes such as over- versus under-personalization
and schedule misalignment versus constraint violation. (RQ2)~An LLM-as-judge pipeline was validated against
human coders, with twelve of fifteen metrics achieving ICC $\geq$ 0.73 and three reaching perfect agreement,
providing a replicable calibration protocol for automated evaluation in domain-specific conversational AI.
(RQ3)~Comparative evaluation across four configurations demonstrated that interaction quality metrics --
particularly over-personalization -- are more discriminating than factual accuracy for differentiating
agentic from non-agentic architectures. Future work will replicate the framework across multiple households
and utility structures, validate system rankings with real residential users, and apply the framework to
emerging conversational HEMS architectures.


\acknowledgments{This material is based upon work partially supported by Salt River Project (SRP) under
grant SRP UA-08 24-25 and the National Science Foundation (NSF) under grant \#2519054.
Any opinions, findings, and conclusions or recommendations expressed in this material are
those of the authors and do not necessarily reflect the views of the NSF or SRP. We would
like to thank our SRP advisor, Victor J.\ Berrios, who provided external evaluation
of our methodological approach.}
\aiuse{During the preparation of this work the author used OpenAI GPT-5.2 and Claude Opus 4.5/4.6 to proofread sentences. The author reviewed and edited all content and takes full responsibility for the publication.}

\appendix
\section{Home Energy Management Assistant (HEMA)}
\label{app:hema}

\begin{figure}
\centering
\includegraphics[width=\columnwidth]{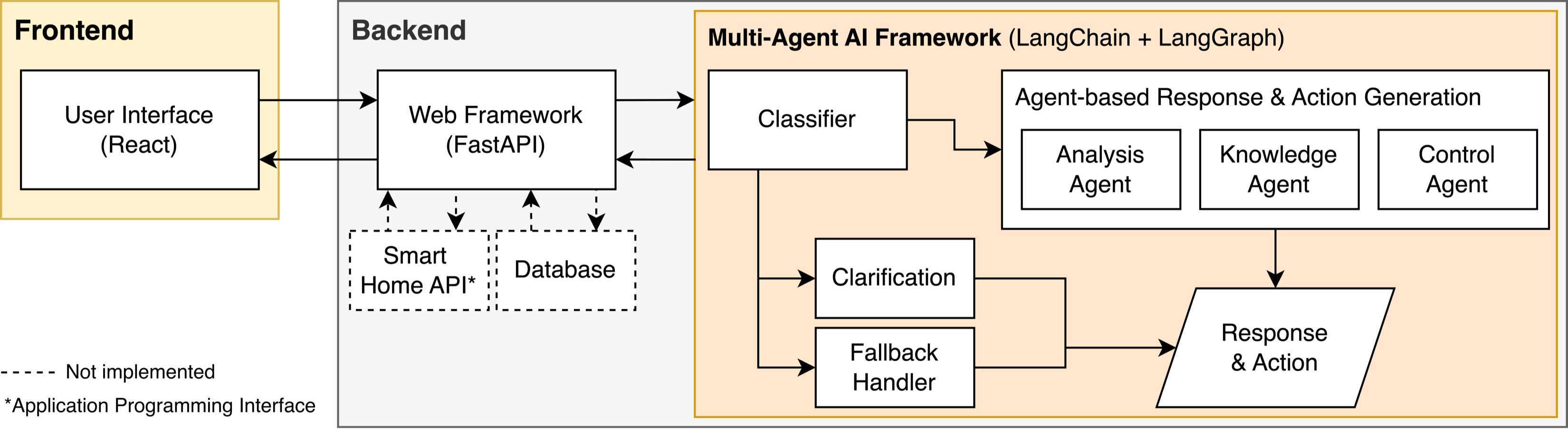}
\caption{HEMA's system architecture}
\label{fig:system_architecture}
\end{figure}

HEMA \cite{jung2026cot} employs a multi-agent architecture (Figure~\ref{fig:system_architecture})
comprising three specialized agents -- Analysis, Knowledge, and Control -- coordinated by a
self-consistency classifier that runs four parallel Chain-of-Thought classification attempts
and aggregates results through majority voting. Each agent follows the ReAct paradigm
\cite{yao2023react}, chaining tool calls with intermediate reasoning before synthesizing
a response. The Analysis agent provides 15 tools for energy data retrieval and statistical
analysis; the Knowledge agent provides 7 tools for RAG over energy documents, weather data,
and built-in concept explanations; and the Control agent provides 9 tools for device
discovery, state monitoring, and command execution following a discovery-first validation
approach. The system is implemented using LangGraph \cite{wang2024langgraph} for deterministic
state-graph routing. Complete tool specifications, implementation details, and source code
are available in \cite{jung2026cot}.

\bibliographystyle{unsrtnat}
\bibliography{refs}

\end{document}